


\def\Re{{\rm Re}}

\def\brk{\hfill\break}
\overfullrule=0pt

\pubnum{6095}
\date{March, 1993}
 \pubtype{T/E}
 \titlepage
 \title{%
Parton-Parton
Elastic Scattering and Rapidity Gaps at Very High Energies\doeack}
\author{Vittorio Del Duca and Wai-Keung Tang}
\SLAC
\abstract{The theory of the perturbative pomeron,
due to Lipatov and collaborators, is used to compute the
probability of observing parton-parton elastic scattering and rapidity
gaps between jets in hadron collisions at very high energies.}
\submit{Physics Letters B}
\endpage
\vfill

The energies of the Tevatron, LHC and SSC colliders open a window
on a range of events in a new
kinematical regime of hadron scattering, the semihard regime, where the
hadron center-of-mass energy $\sqrt{s}$ is much larger than the typical
parton transverse momentum $m$. This complicates the experimental
analysis of such events, since it requires jet detection in a
large rapidity interval\REF\fad{J.D. Byorken, \sl Int. J. Mod. Phys.
\bf A7 \rm (1992) 4189.}\refend.
It also makes harder the calculation of jet cross sections, since
it involves large ratios of kinematical invariants which may appear
either in the short distance cross section \REF\MN{A. H. Mueller and
H. Navelet, \sl Nucl. Phys. \bf B282 \rm (1987) 727.}\refend\ or in
the evolution of the parton structure functions \REF\smallpart{J.C.
Collins and R.K. Ellis, \sl Nucl. Phys. \bf B360 \rm (1991) 3;
\brk S. Catani, M. Ciafaloni and F. Hautmann, \sl Nucl. Phys. \bf B366
\rm (1991) 135.}\refend, thus requiring the introduction of
techniques to resum infinite classes of Feynman diagrams.
Nonetheless the study of jet cross
sections at very high energies is very interesting, since it may
provide new insights on the dynamics of QCD processes and offer
signatures of novel physical events\REF\BJ{J. D. Bjorken,
\sl Phys. Rev. \bf D47 \rm (1992) 101.}\refend.

In     order    to    obtain    quantitative
predictions   of jet production in the semihard regime
and separate it from the  uncertainty
involving  the small  $x$  dependence  of parton
distributions\refmark\smallpart, Mueller and
Navelet \refmark\MN\ proposed to measure the two-jet
inclusive cross section in hadron
collisions by tagging two jets at a large rapidity interval $y$ and with
transverse  momentum  of order $m$.
These tagging jets are produced in a nearly forward scattering of gluons or
quarks with  large center-of-mass energy $\sqrt{\hat s}$.
\REF\BFKLone{L. N. Lipatov, \sl Sov. J. Nucl. Phys. \bf 23 \rm
(1976) 338.}
\REF\BFKLtwo{
E. A. Kuraev, L. N. Lipatov,
and V. S. Fadin, \sl Sov. Phys. JETP \bf 44 \rm (1976) 443, \bf
 45 \rm (1977) 199.}
\REF\BFKLthree{Ya. Ya.
Balitsky and L. N. Lipatov, \sl Sov. J. Nucl. Phys.
\bf 28 \rm (1978) 822; \brk
L. N. Lipatov, \sl Sov. Phys. JETP \bf 63 \rm (1986) 904.}
\REF\LP{L. N. Lipatov, in \sl Perturbative QCD, \rm ed. A.H. Mueller (World
Scientific, Singapore, 1989).}
Lipatov and collaborators\refmark{\BFKLone-\LP} (BFKL) have shown that,
in this regime, the rapidity interval $y=\ln({\hat s}/m^{2})$
between the scattered partons is filled in by the radiation of additional
gluons, roughly uniformly spaced in rapidity, all with transverse
momenta of order $m$.
 \FIG\BFKLg{BFKL resummation: (a) emission of gluons; (b) construction
    of the octet amplitude; (c) construction of the singlet amplitude.}
The BFKL theory systematically corrects the lowest-order QCD result
 by summing the leading logarithms of $\hat s$.
This is done in three stages, as shown in Fig. \BFKLg.
First, one simplifies the lowest-order QCD diagrams for multigluon
production, shown in Fig. \BFKLg(a), for the case in which the emitted
gluons are widely separated in rapidity. The gluon emission vertex is
replaced by a non-local gauge-invariant effective vertex
\refmark\BFKLone. Next, one sums the leading
corrections to the forward
amplitude with  color octet exchange in the $t$-channel, as
shown in Fig.  \BFKLg(b).  The result has the form of a Regge pole
with an infrared-sensitive trajectory\refmark\BFKLtwo.
Finally, one uses this resummed, effective
gluon exchange to compute the elastic amplitude in the Regge limit
$\hat s \gg -t$ with color singlet exchange in the $t$-channel, as shown in
fig.\BFKLg(c). This is known as the BFKL pomeron\refmark\BFKLthree.
The imaginary part of the forward amplitude is the
parton-parton total cross section.
To leading order in rapidity, the parton-parton total
cross section and the related
2-jet inclusive cross section exhibit the energy
dependence $\exp[(\alpha_{P}-1)y]$ with
$$\alpha_{P}=1+4\ln2{\alpha_s C_A \over \pi}, \eqn\alph$$
where $C_A = N_c = 3$ is the number of colors in QCD.

\FIG\BFKLcross{BFKL Pomeron: (a) in minijet production; (b) in high
energy elastic scattering.}
A large total cross section $\hat \sigma_{total}$ for parton-parton
scattering implies also a large elastic scattering cross section,
since the total cross section is related, through the optical theorem,
to the elastic scattering amplitude with color singlet exchange in the
$t$ channel. Thus both elastic scattering and jet
production can be described by exchanging one BFKL pomeron Fig.
\BFKLcross(a,b). This process has been studied
by Mueller and one of the authors\REF\MT{A. H. Mueller and W.-K. Tang,
\sl Phys. Lett. \bf B284 \rm (1992) 123.}\refend. It is a higher order
($\alpha_s^4$) process but with energy dependence $\exp[2(\alpha_{P}-1)y]$.
However, it leads to a final state which, at the parton level, contains
two jets with a rapidity gap in gluon production between
them\refmark\BJ. Some fraction of these states may produce the
dramatic experimental signatures of a large rapidity gap in secondary
particle production. To understand the relation between rapidity gaps
in hard-gluon and hadron production, we must discuss the potential
backgrounds to these signals at the parton and hadron level.

To analize the parton-level background, assume that we cannot
detect partons with transverse momentum smaller
than a fixed parameter $\mu$. In this case, there is an additional
contribution to elastic scattering
from color octet exchange in the $t$
channel. This is shown in fig.\BFKLg(b) where a reggeized gluon is
exchanged. The reggeized gluon contains all the leading virtual
radiative corrections and it has the form of a Sudakov form factor.
The parameter $\mu$ fixes the scale below which soft gluon radiation is
allowed. As $\mu \rightarrow 0$, the contribution of the color octet
exchange vanishes, since it is impossible to have scattering with
exchange of a gluon, without allowing for the emission of soft gluon
radiation.

In order to use perturbative QCD, the parameter $\mu$ must be larger
than $\Lambda_{QCD}$. Thus we have two options:
first, we can consider $m \gg \mu \gg \Lambda_{QCD}$, that is, we define a
rapidity gap to be present if there are no jets between the tagging
jets. We will call this case $quasi\, elastic\,
scattering$, since it allows gluon radiation below the scale $\mu$.
The ratio $R$ of the quasi-elastic to the total
cross section is given by
$$ R(\mu) = {\sigma_{singlet} + \sigma_{octet} \over \sigma_{tot}}. \eqn
   \ratio$$
where all the cross sections in \ratio\ have been convoluted with the
appropriate parton distributions.
\Ref\GLM{E. Gotsman, E.M. Levin and U. Maor, \sl TAUP-2030-93 \rm
preprint.}
Alternatively, we can consider
$\mu = O(\Lambda_{QCD})$. Then at the parton level the color octet
exchange is strongly suppressed, and only the color singlet exchange
contributes to the cross section for producing rapidity gaps.

At the hadron level, there is an additional consideration. Accompanying
any hard QCD reaction in hadron-hadron collisions, there are spectator
partons. These may produce hadrons across the rapidity interval,
spoiling the rapidity gap. Thus in
order to compute the cross section for producing a rapidity gap at the
hadron level, we need a non-perturbative model which describes the hadron
interaction and estimates the survival of the rapidity gap in the presence
of soft spectator interactions\refmark\BJ.
The rapidity-gap survival probability $<S^2>$ is defined
as the probability that in a scattering event no other interaction
occurs beside the hard collision of interest. This probability is most
readily estimated as an average over the hadron-hadron impact parameter
$B$\refmark\BJ:
$$ <S^2> = {\int d^2B f(B) S^2(B) \over \int d^2B f(B)}, \eqn\gap$$
where $S^2(B)$ is the probability that the colliding hadrons
do not interact inelastically, and
$f(B)$ is the cross section for the hard collision of interest.
Different estimates for $<S^2>$ in hadron collisions,
based on a variety of phenomenological models, are presented in
ref.\GLM, where $<S^2>$ is estimated to be between 0.05 and 0.2.
$<S^2>$ is expected to depend on the hadron-hadron center of mass
energy, but only weakly on the size of the rapidity gap.
Then to obtain the probability of a scattering
event with a large rapidity gap at the hadron level, we must compute
the ratio $R$ at $\mu = 0$, that is, using only the singlet elastic
cross section, and multiply it
by the survival probability $<S^2>$:
$$ R_{gap} = <S^2> R(\mu = 0). \eqn\rgap$$
In this paper, we will present the asymptotic form of $R(\mu)$, and a
numerical estimate for this quantity for rapidity intervals $y<15$.

For very large rapidity intervals, the one pomeron exchange
approximation violates unitarity since the
elastic cross section becomes larger than the
total cross section. Our estimate
indicates that the unitarity bound for the one pomeron exchange
approximation is at about 24 units of rapidity, when the transverse
momentum of the tagging jets $m$ is about 20 GeV. This is far beyond the
maximum rapidity interval available at the SSC. At lower jet
transverse momenta or for larger rapidity intervals,
$R$ must be unitarized by
 the contribution from multiple pomeron exchange.

\noindent{\bf Total Cross Section for Tagging Jets}

\REF\DPT{V. Del Duca, M. E. Peskin and W.-K. Tang,
      \sl SLAC-PUB-\bf 6065 \rm preprint, to appear in \sl Phys. Lett. B\rm.}
   The total cross section for two tagging jets has been studied by
Mueller and Navelet\refmark\MN, who used the BFKL analysis
of the asymptotic parton-parton total cross section to derive the 2-jet
inclusive cross section.
Following the exposition of ref. \DPT, we consider the scattering of two
hadrons
of
momenta $p_A$ and $p_B$ in the center-of-mass frame, with the $z$ axis
along the beam momenta, and we imagine to  tag two jets at the extremes
of the Lego plot, with the rapidity interval between them filled
with jets. We will call these extra jets $minijets$. The tagging jets
can be characterized by their transverse momenta
and by their longitudinal fractions $x_A$, $x_B$ with
respect to their parent hadrons.  It is simplest to consider the
cross section for producing two tagging jets with transverse momenta
greater than a minimum value $m$.  Then
$$\eqalign{
 {d\sigma_{tot} \over dx_A dx_B}&(AB\rightarrow j(x_A) j(x_B) + X) \cr
&  = \prod_{i=A,B}
   \biggl[G(x_i,m^2) + 4/9 \sum_f [Q_f(x_i,m^2) +
   \bar Q_f(x_i,m^2)]\biggr]  \cdot
   \hat\sigma_{tot}(\hat s),  \cr}
\eqn\totaltagrate$$
where $\hat s = 2p_A\cdot p_B x_A x_B$ is the parton-parton squared
center-of-mass energy, and
 $\hat\sigma_{tot}$ is the BFKL total cross section for
gluon-gluon
scattering within an impact distance of size $1/m$.   Eq. \totaltagrate\
includes the effects of quarks using the observation of Combridge
and Maxwell that, in a process with large rapidity intervals, the
leading contribution to any scattering process comes from gluon
exchange in the crossed channel\Ref\CM {B. L. Combridge
and C. J. Maxwell,
\sl Nucl. Phys. \bf B239 \rm (1984) 429.}. The factor 4/9 in
eq.\totaltagrate\ is the ratio of the Casimir operators $C_F/C_A$,
with $C_F=(N_c^2-1)/2N_c$.
The values of $x_A$, $x_B$ should be taken to be sufficiently large that the
parton
distributions satisfy ordinary DGLAP (Dokshiter, Gribov, Lipatov,
Altarelli and Parisi) evolution;
the BFKL theory adds additional complications in the evolution of the
parton structure functions when these fractions
become small\refmark\smallpart.

The core of eq. \totaltagrate\ is the BFKL function
$\hat\sigma_{tot}(\hat s)$, which is given by

 $$ \hat\sigma_{tot}(\hat s)= {8 \over N_c^2 -1} \Bigl( {\alpha_s
 C_A \over \pi} \Bigr)^2 {\pi^3 \over 2 m^2}
   F_{tot}(y). \eqn\sigmatot$$
$F_{tot}(y)$ is a dimensionless function
which we will discuss below.  The strong coupling constant is evaluated
at a scale $m^2$; the running of $\alpha_s$ is subleading in the BFKL
theory.
  $F_{tot}$ is normalized to $1$  at the lowest order of QCD
for the gluon-gluon scattering cross section, integrated
over the transverse momenta of the tagging jets with
$|k_{i\perp}| \geq m$, in the limit
$\hat s>>m^2$. Since $F_{tot}$ is always larger than 1, it can be
viewed through the optical theorem as the
enhancement factor due to pomeron exchange.
$F_{tot}$ can be calculated from the
solution of an integral equation for the elastic scattering amplitude,
with color singlet exchange, constructed by
BFKL\refmark{\BFKLtwo - \LP}. It is given by\refmark\MN

$$ F_{tot}(y) =  \int_{-\infty}^{\infty}
   {d\nu\over 2 \pi} {1 \over \nu^2 + 1/4} e^{\omega(\nu)y},
   \eqn\ffinal$$
with
$$\omega(\nu) = \omega(0,\nu) \eqn\ultima$$
the value at $n=0$ of the eigenvalue of the
integral equation
$$ \omega(n,\nu) = {2 \alpha_s C_A \over\pi}\bigl[ \psi(1) - \Re\psi
   ({|n|+1\over 2} +i\nu) \bigr],
   \eqn\omegeval$$
and $\psi(z)$ the standard logarithmic derivative of the Gamma function.
The exponential growth of
$F_{tot}(y)$ with the rapidity interval is associated with minijet
production.
 Near $\nu = 0$, $\omega(\nu)$ has the expansion

$$   \omega(\nu)  =  A  -  B\nu^2 + \cdots , \eqn\omegaexpand$$
with
$$     A =  {4\alpha_s C_A\over \pi}\log 2
   , \qquad    B = {14\alpha_s C_A\over \pi} \zeta(3).
\eqn\AandBforms$$

Using \omegaexpand\ to expand about the saddle point
at $\nu = 0$, we can see that $F_{tot}(y)$
 has the asymptotic behavior

$$    F_{tot}(y) \sim {e^{(4\log 2) z}\over \sqrt{{7\over 2}\zeta(3) \pi z} },
\qquad {\rm with }\   z = {\alpha_s C_A\over \pi} y .
\eqn\asymptotsZ$$

Mueller and Navelet showed that this asymptotic form is
an accurate representation of eq.\ffinal\ for $z> 0.2$.

\noindent{\bf Elastic Cross Section for Tagging Jets}

The high energy elastic scattering cross section with color singlet exchange in
the $t$ channel has been studied in ref. \MT. Then it is straightforward to
write down the elastic cross section for tagging jets, with singlet
exchange, as

$$\eqalign{
 {d\sigma_{singlet} \over dx_A dx_B}&(AB\rightarrow j(x_A) j(x_B) ) \cr
&  = \prod_{i=A,B}
   \biggl[G(x_i,m^2) + (4/9)^2 \sum_f [Q_f(x_i,m^2) +
   \bar Q_f(x_i,m^2)]\biggr]  \cdot
   \hat\sigma_{singlet}(\hat s).  \cr}
\eqn\elasticsinglet$$
The BFKL function $\hat\sigma_{singlet}(\hat s)$ is the gluon-gluon
elastic scattering cross section
in the color singlet channel with the tagging
jets collimated  and with minimum jet transverse momentum $m$.
Since two reggeized gluons are involved in the color singlet
exchange in the $t$ channel, in keeping into account the possibility
that the scattering is initiated by quarks we obtain the suppression
factor $(C_{F}/C_{A})^2$.
$\hat\sigma_{singlet}(\hat s)$ is given by

$$ \hat\sigma_{singlet}(\hat s) = \Bigl( {8 \over N_c^2 -1} \Bigr)^2
   \Bigl( {\alpha_{s} C_A \over \pi} \Bigr)^4
   {\pi^5 \over 4 m^2} F_{singlet}(y), \eqn\elasticinteg$$
with

$$ F_{singlet} =  {m^2 \over 16\pi^3} \int_{m^2} d^2k_{A\perp}
\;\;\left| \int
d^2q_{\perp}d^2q'_{\perp}f^{k_{A\perp}}(q_{\perp},q'_{\perp},y)\right|^2.
\eqn\enhancement$$

The amplitude $f^{k_{A\perp}}(q_{\perp},q'_{\perp},y)$ represents the
propagation of a BFKL pomeron in transverse momentum plane
(fig.\BFKLcross(b)), with $t = -k_{A,\perp}^2$.
The normalization of $f^{k_{A\perp}}(q_{\perp},q'_{\perp},y)$ is as in
ref. \LP\ so that at
lowest order

$$f^{k_{A\perp}}(q_{\perp},q'_{\perp},y)= {\delta^2(q_{\perp}-q'_{\perp}) \over
q_{\perp}^2 (q'_{\perp}-k_{A\perp})^2}. \eqn\normalization$$

Lipatov\refmark{\BFKLthree,\LP} gives an integral representation of
$f^{k_{A\perp}}(q_{\perp},q'_{\perp},y)$ in impact parameter space, for
the scattering of color neutral objects. The extension to parton-parton
scattering has been done in ref.\MT. The integral of
$f^{k_{A\perp}}(q_{\perp},q'_{\perp},y)$, over its
transverse momenta, is\refmark\MT

$$ \int
d^2q_{\perp}d^2q'_{\perp}f^{k_{A\perp}}(q_{\perp},q'_{\perp},y) =
{4\over k_{A\perp}^2} \int d\nu {\nu^2\over (\nu^2+{1\over 4})^2}
\exp[\omega(\nu)y], \eqn\amplitude$$
where only the leading
term $n=0$, for which $\omega(\nu)$ is given by \ultima\ and \omegeval,
has been kept.
As Mueller and Tang have shown, the integration over the transverse
momenta in \amplitude\ is not infrared sensitive, i.e. the infrared
divergences in $q_{\perp}$, which are present in the lowest order contribution
\normalization, disappear when we consider the infinite resummation
\amplitude\ of the leading logarithmic corrections.
It is an interesting result that the pomeron trajectory has
no explicit dependence on the momentum transfer. The dependence on $t$
is only through the coupling constant.

By substituting \amplitude\ into \enhancement\ we calculate $F_{singlet}$.
The integral over $k_{A\perp}$ is singular and depends on the cutoff $m$;
this gives us the same factor $m^{-2}$ as in the total cross section. The
enhancement factor becomes
$$ F_{singlet}(y)= 4 \left( \int {d\nu \over 2\pi} {\nu^2\over (\nu^2+{1\over
4})^2}
\exp[\omega(\nu)y] \right)^2. \eqn\enhancementfinal$$
$F_{singlet}$ is normalized in order to have  $F_{singlet} =1$ at $y=0$.
Notice, though, that $F_{singlet}$ at $y=0$ from \enhancementfinal\
does not correspond to gluon-gluon elastic scattering at the lowest
order of QCD, which is instead correctly given substituting
\normalization\ into \enhancement.
Using the expansion of $\omega(\nu)$ around the saddle point $\nu=0$,
we obtain the asymptotics of $F_{singlet}$
$$    F_{singlet}(y) \sim  \left(\pi {e^{(4\log 2) z}\over \bigl({{7\over 2}
\zeta(3) \pi z}\bigr)^{3/2} }\right)^2. \eqn\enhancementasy$$
where $z$ is defined in \asymptotsZ.

The background to the color singlet exchange comes
from the exchange of a reggeized gluon. This contribution is
$$\eqalign{
 {d\sigma_{octet} \over dx_A dx_B}&(AB\rightarrow j(x_A) j(x_B) ) \cr
&  = \prod_{i=A,B}
   \biggl[G(x_i,m^2) + 4/9 \sum_f [Q_f(x_i,m^2) +
   \bar Q_f(x_i,m^2)]\biggr]  \cdot
   \hat\sigma_{octet}(\hat s,\mu),  \cr}
\eqn\elasticoctet$$
$\hat\sigma_{octet}(\hat s,\mu)$ is the gluon-gluon
elastic scattering cross section in the color octet channel, with the
tagging jets collimated and with minimum transverse momentum $m$.
We may write $\hat\sigma_{octet}(\hat s,\mu)$ as\refmark\MT

$$ \hat\sigma_{octet}(\hat s,\mu)= {8 \over N_c^2-1} \Bigl({\alpha_{s}
   C_A \over
\pi}\Bigr)^2 {\pi^3 \over 2 m^2} F_{octet}(y), \eqn\octetinteg$$
where $F_{octet}$ is the solution of the BFKL integral equation for
elastic scattering with color octet exchange\refmark\BFKLtwo
$$ F_{octet}=\exp[2(\alpha_{g}-1)y]. \eqn\octetsuppression$$
The reggeized gluon trajectory is
$$ \alpha_{g} = 1 - {\alpha C_A \over  \pi} \log \left( {m\over \mu}
\right), \eqn\gluontrajectory$$
for $m^2/\mu^2 \gg 1$. From the expression of the gluon trajectory, we can
see that $\alpha_{g}-1 < 0$ and hence $F_{octet}$ is always less than
one. As $\mu \rightarrow 0$, or the rapidity interval $y$ becomes
large, $\hat\sigma_{octet}(\hat s,\mu)$ vanishes.

\noindent{\bf The Ratio $R(\mu)$}

$R(\mu)$ is the probability of having elastic scattering at the
parton level, as defined in \ratio. By substituting \elasticoctet,
\elasticsinglet\ and
\totaltagrate\ into the definition of $R(\mu)$, we obtain
$$ R(x_{A},x_{B}, \mu) = {\hat\sigma_{octet}\over \hat\sigma_{tot}}
                         + w_{f}(x_A) w_f(x_B)
                         {\hat\sigma_{singlet}\over\hat\sigma_{tot}}
                         \eqn\ratiosigma$$
where $w_{f}$ is a weight of parton distributions,
$$    w_{f}(x)= {
   \biggl[G(x,m^2) + (4/9)^2 \sum_f [Q_f(x,m^2) +
   \bar Q_f(x,m^2)]\biggr]  \over
   \biggl[G(x,m^2) + (4/9) \sum_f [Q_f(x,m^2) +
   \bar Q_f(x,m^2)]\biggr]} \eqn\weight$$
and
$$ \eqalign{{\hat\sigma_{singlet}\over\hat\sigma_{tot}} = \;\;&
{8 \over N_c^2 - 1} \Bigl({\alpha_{s} C_A
\over \pi}\Bigr)^2 {\pi^2\over 2} {F_{singlet}\over F_{tot}} \cr
 {\hat\sigma_{octet}\over \hat\sigma_{tot}} =\;\;&  {F_{octet}\over
 F_{tot}}. \cr} \eqn\sigmaratio$$
 In the asymptotic regime, where the rapidity $y$ is large,
$$ \eqalign{{F_{singlet}\over F_{tot}} \sim \;\;& \pi^2
 {e^{(4\log 2) z}\over ({{7\over 2}\zeta(3) \pi z})^{5/2} }  \cr
 \noalign{\smallskip}
 {F_{octet}\over F_{tot}} \sim \;\;&
 e^{-(4\log 2 + 2\log (m/\mu)) z} \sqrt{{7\over 2}\zeta(3) \pi z}  \cr}
 \eqn\Fasymptotics$$
 and so the octet contribution can be neglected in the asymptotic
 regime. The asymptotic expressions can be used to estimate the
range of  validity of the one pomeron exchange approximation.
 Taking $\alpha_{s}\sim 0.15$, which is the typical value
 for tagging jets of transverse momentum $m = 20 GeV$, the violation
 of the unitarity bound happens at a rapidity interval $y \simeq 24$.

 \noindent{\bf Numerical Evaluation}

 We now turn to a numerical analysis of $R(\mu)$. The formulae
 \ffinal, \enhancementfinal, \octetsuppression\ and
 \ratiosigma\ are the appropriate starting point for a numerical
 evaluation of the ratio $R(\mu)$. We scale the running coupling
 constant from $\alpha_{s}(m(Z))=0.12$ using the 1-loop evolution with 5
 flavors, and take the longitudinal fractions $x_{A}$,
 $x_B$ of the tagging jets to be $0.1$.

 \FIG\ratio{Ratio $R(\mu)$ as a function of the rapidity $y$ of the
 tagging jets. We choose $x_A$ and $x_B$ to be $0.1$. The dotted line is
 the color singlet contribution. The solid line includes the contribution
 from both singlet and octet. We label the curves according to the
 minimum jet transverse momentum. (a) $\mu = 1\hbox{GeV}$; (b) $\mu =
 5\hbox{GeV}$.}

In Fig. \ratio, we show the result for $m=10-40 \hbox{GeV}$. Even
though we present the result from $y = 0$, the plots cannot be trusted
in the small rapidity regime since the leading logarithmic approximation
is no longer valid there. The dotted line is the contribution from color
singlet exchange only, and thus corresponds to the choice $\mu = 0$, while
the solid line includes both singlet and octet exchanges. In Fig. \ratio(a),
$\mu$ is set to  1GeV. For $y > 7$,
only the color singlet contributes since the octet is highly suppressed.  In
Fig. \ratio(b), $\mu$ is set to 5GeV. The suppression of the octet
occurs now for $y > 10$.
The two figures show the same trend in the asymptotic regime
($y>10$) since the singlet exchange is independent of the choice of
$\mu$.

The value of $R(\mu = 0)$, given by the dotted line in Fig. \ratio,
multiplied by the survival propability $<S^2>$ \gap, gives the
probability of having a collision with a large rapidity gap \rgap.
Bjorken\refmark\BJ\ has estimated $<S^2>$ to be $\simeq 0.1$, and the
authors of ref.\GLM\ have given estimates in the range of 0.05 to
0.2. Thus, we expect that, at the Tevatron, a few tenths percent of
events with tagging jets will show rapidity gaps in hadron production.
The probability of finding a gap increases with the rapidity interval
between the tagging jets, as indicated in the figure. Though $<S^2>$
depends on the collider center of mass energy, it should depend only
weakly on rapidity. Thus, we expect that the probability of observing a
rapidity gap in secondary particle production between tagging jets
should rise proportionately to the dashed curves in Fig. 3.

Since all of the analysis above is in the leading logarithmic approximation,
there is ambiguity in the choice of the proper scale in rapidity for
which this analysis is valid, and so the exact value of the
normalization and thus of $R(\mu)$ cannot be determined precisely.
We need a next-to-leading order calculation, which is not available
yet, to have a definite
quantitative prediction. However, the slope of the curves in the
asymptotic regime is free from this scale uncertainty and thus the
experimental measurement of the ratio $R(\mu)/R_{gap}$ in the large
rapidity-gap regime should give
us an unambiguous determination of the survival probability $<S^2>$.
It will be interesting to see if $<S^2>$ depends on the width of the rapidity
gap between the tagging jets, when the center-of-mass energy of the
colliding hadrons is kept fixed.

\noindent{\bf Acknowledgements}

We are very grateful to bj. Bjorken and Michael Peskin for encouraging
us to investigate this problem, and for many stimulating discussions
and suggestions, and for constructive criticisms while reading this
manuscript.

\endpage

\refout
\endpage

\figout

\end
\bye